\documentclass[twocolumn]{revtex4}

\makeatletter

\usepackage{graphicx}
\usepackage{amsmath}

\makeatother

\begin{document}

\title{Swimming Efficiency of Bacterium Escherichia Coli}

\author{Suddhashil Chattopadhyay$^{1}$, Radu Moldovan$^{1}$,
Chuck Yeung$^{2}$, and X.L. Wu$^{1}$}
\email{xlwu@pitt.edu}

\affiliation{$^{1}$Department of Physics and Astronomy, University of Pittsburgh,
Pittsburgh, PA 15260}

\affiliation{$^{2}$School of Science, Pennsylvania State University at Erie,
The Behrend College, Erie, PA 16563}

\date{Dec 22, 2005}

\begin{abstract}
We use measurements of swimming bacteria in an optical trap to determine
fundamental properties of bacterial propulsion. In particular, we
determine the propulsion matrix, which relates the angular velocity
of the flagellum to the torques and forces propelling the bacterium.
From the propulsion matrix, dynamical properties such as forces, torques,
swimming speed and power can be obtained by measuring the angular
velocity of the motor. We find significant heterogeneities among different
individuals even though all bacteria started from a single colony.
The propulsive efficiency, defined as the ratio of the propulsive
power output to the rotary power input provided by the motors, is
found to be 0.2\%. 
\end{abstract}

\maketitle

Bacteria swim by rotating helical propellers called flagella. In the
case of Escherichia coli (E. coli), each flagellum is several microns
in length, 20 nm in diameter and four to five of them organize into
a bundle. The flagella are driven at their bases by reversible rotary
engines that turn at a frequency of approximately $100~Hz$. Existing
experiments show that these molecular engines are Poisson stepping
motors consisting of several hundred steps per revolution \cite{Berg:PoissonMotor,berry:steps}.
However many essential properties of bacterial swimming have not been
measured, particularly in intact cells. For example, what is the relation
between the angular velocity of the propellers and the force propelling
the bacteria forward? What fraction of the flagellar motor power is
converted into translational motion? What variability is there in
the swimming apparatus from cell to cell? Some of these fundamental
questions have been addressed in theoretical and numerical work \cite{taylor:wavingtails,Lighthill:BiofluidDynamics,childress},
however, direct measurements of intact cells with functional motors
and flagella are limited \cite{berg:proton,berg:fluorescent,magariyama}.

Herein, we report an investigation of the fundamental swimming properties
of E. coli using optical tweezers and an imposed external flow. We
measure the force required to hold the bacterium, and the angular
velocities of the flagellar bundle and of the cell body as a function
of the flow velocity. The propulsion matrix, which relates the translational
and angular velocity of the flagella to the forces and torques propelling
the bacterium, can thus be determined one bacterium at a time. Our
measurements show that although the population averaged matrix elements
are in reasonably good agreement with the resistive force theory for
helical propellers \cite{Lighthill:BiofluidDynamics}, there is a
large variability among bacteria from a single colony. The propulsion
matrix also allows us to determine the propulsive efficiency $\varepsilon$,
defined as the ratio of the propulsive power output to the rotary
power input, to be $0.2\%.$ This is consistent with experiments on
helical propellers \cite{purcell} and close to the maximum efficiency
for the given cell-body and shape of the flagella. It is smaller than
the 1 or 2\% predicted theoretically for simple shapes such as a corkscrew
\cite{childress}. Our experimental technique is versatile and can
be used to make comparative studies of mutants strains of the same
species or of different micro-organisms. Such measurements can shed
new light on how this remarkable ability to swim evolves among different
bacterial species.

\begin{figure}

\centerline{\includegraphics[width=8.5cm]{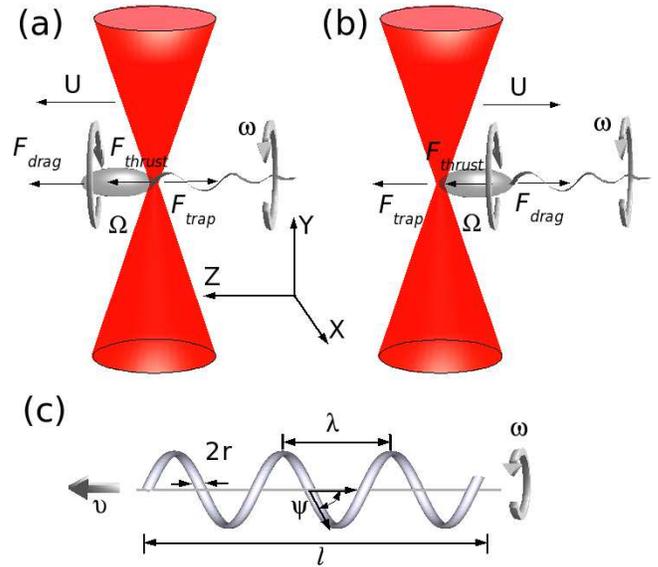}}

\caption{Two different trapping configurations are possible. (a) The bacterium
can be trapped at the tail of the cell body in the presence of an
imposed flow $U$. The trapping is stable for $U\geq0$. (b) The bacterium
can also be trapped at the head of the cell body for $U$ between
$-40~\mu m/s$ and $-100~\mu m/s$. The forces and velocities are
positive if they are along $+Z$. The rotations are defined by the
right-hand rule such that $\omega<0$ and $\Omega>0$ as depicted.
(c) A schematic of a helical flagellum: $\ell$ is the length, $2r$
is the diameter of the filament, $\Psi$ is the pitch angle of the
helix relative to the swimming axis, and $\lambda$ is the pitch.
\label{cap:experimentalsetup}}
\end{figure}

An important feature of bacterial swimming is that at very low Reynolds
numbers ($Re\simeq10^{-4}$), the fluid motion is governed by Stokes
flow and nonlinearities in the full hydrodynamic equation are irrelevant.
Despite this simplifying feature, the problem remains theoretically
difficult due to complicated time-dependent boundary conditions. Theoretical
studies, therefore, usually assume that the flagella have very simple
geometries such as an infinite sheet \cite{taylor:wavingtails} or
a helical coil \cite{Lighthill:BiofluidDynamics,childress}. A second
approach is not to take into account specific geometries but to consider
general relations appropriate in the low Reynolds number limit \cite{purcell}.
In this regime, the torque $N_{fl}$ acting on the propeller and the
thrust force $F_{thrust}$ generated by the propeller are linearly
related to the propeller's angular velocity $\omega$ and the translational
velocity $v$ (relative to the background fluid): \begin{subequations}\label{eq:propulsion}
\begin{eqnarray}
-F_{thrust} & = & Av-B\omega \\
N_{fl} & = & -Bv+D\omega.
\end{eqnarray}
\end{subequations} The above equation can be expressed in terms of the symmetric propulsion
matrix $P=\left[\begin{array}{cc}
A & -B\\
-B & D\end{array}\right]$, also known as the resistance matrix \cite{happel:lowreynolds}.
Choosing the coordinate system in Fig. \ref{cap:experimentalsetup},
$F_{thrust}$ and $v$ are positive if directed toward the head of
the cell while the sign of $\omega$ and $N_{fl}$ obeys the right-hand
rule, i.e., the flagella is a left-handed helix. Based on this coordinate
system, the coefficients $A$, $B$, and $D$ are positive, proportional
to fluid viscosity $\eta$, and depend on the shape and size of the
propeller. The basic physics is that in the absence of an applied
torque, a translating propeller under the influence of an external
force must rotate, and in the absence of an applied force, a rotating
propeller under the influence of an external torque must translate
\cite{purcell}. The above formulation is applicable to propellers
of any shape and size. However, for the special case of a helical
coil, the matrix elements can be derived from resistive force theory
\cite{Lighthill:BiofluidDynamics}: \begin{subequations} \label{eq:coefficient}
\begin{eqnarray}
A & = & K_{n}\ell~\frac{(1-\beta(1-\gamma_{k}))}{\beta^{1/2}},\label{eq:A-coefficient} \\
B & = & K_{n}\ell~\frac{\lambda~(1-\beta)(1-\gamma_{k})}{2\pi\beta^{1/2}},\label{eq:B-coefficient}\\
D & = & K_{n}\ell~\frac{\lambda^{2}~(1-\beta)}{4\pi^{2}~\beta^{1/2}}~\left(1+\gamma_{k}\frac{(1-\beta)}{\beta}\right),\label{eq:D-coefficient}\end{eqnarray}
\end{subequations} where $\ell$ is the length of the coil, and $\beta=\cos^{2}\Psi$
with $\Psi$ being the pitch angle relative to the swimming axis (see
Fig.\ \ref{cap:experimentalsetup}c). The quantity $\gamma_{k}$
is the ratio of the tangential viscous coefficient $K_{t}=4\pi\eta/(2\ln(0.18\lambda/r)-1)$
to the perpendicular viscous coefficient $K_{n}=8\pi\eta/(2\ln(0.18\lambda/r)+1)$,
where $\lambda$ is the pitch and $r$ is the radius of the coil filament.
For a smooth coil, Lighthill \cite{Lighthill:BiofluidDynamics} predicts
that $\gamma_{k}=K_{t}/K_{n}\approx0.7$. As can be seen, the helix
loses its ability to propel, if $\gamma_{k}\rightarrow1$, $\Psi\rightarrow0~(\beta\rightarrow1)$
or $\Psi\rightarrow\frac{\pi}{2}~(\lambda\rightarrow0)$ as expected.

To complete the model of the swimming bacterium, we need the propulsion
matrix $P_{0}$ for the cell body. Unlike $P$ for the flagellum,
$P_{0}$ is diagonal ($B_{0}=0$) since the cell body cannot propel
itself. The non-viscous force on the cell body consists of two parts,
the trapping force $F_{trap}$ due to the optical tweezer holding
the bacteria and the thrust $F_{thrust}$ generated by the flagella.
The sum of these forces should balance the viscous force $A_{o}v$
acting on the cell body. Likewise, the non-viscous torque acting on
the cell body $-N_{fl}$ should be balanced by the viscous rotational
drag. This gives: \begin{subequations}\label{eq:propelcellbody} \begin{eqnarray}
F_{trap}+F_{thrust}=A_{0}v,\\
D_{0}\Omega=-N_{fl}, \end{eqnarray} \end{subequations} where $\Omega$ is the angular velocity of the cell body. We treat
the cell body as a prolate with minor semi-axis $a$ and major semi-axis
$b$ so that the linear and rotational drag coefficients are $A_{0}=4\pi\eta~b/(\ln(\frac{2b}{a})-\frac{1}{2})$
and $D_{0}=16\pi\eta a^{2}b/3$ \cite{berg:RandomWalk}. The optical
trapping force is harmonic $F_{trap}(z)=-k(z-z_{0})$, where $k$
is the spring constant and $z-z_{0}$ is the displacement from the
center of the trap \cite{bustamante:trap,Neuman:opticalreview}. Since
the bacteria held by the optical tweezer, its net velocity in the
lab frame is zero ($v'=v+U\simeq0$), and the relative velocity $v$
is opposite to the external flow $U$. Substituting $v=-U$ into Eqs.
\textbf{\ref{eq:propulsion}} and \textbf{\ref{eq:propelcellbody}}
gives: \begin{subequations}
\begin{eqnarray}
k(z-z_{0}) & = & (A+A_{0})U+B\omega,\label{eq:basica}\\
D_{0}\Omega & = & -BU-D\omega.\label{eq:basicb}
\end{eqnarray}
\end{subequations} This set of equations will be used below to analyze our data.

We used a non-tumbling strain of bacteria (RP5231) in our measurements.
We were delighted to find that such a bacterium swimming a few microns
above a glass surface could be stably trapped, along its swimming
direction, by the optical tweezer \cite{com7:}. The bacterium can
then be manipulated by an imposed external flow. Figure \ref{cap:experimentalsetup}
illustrates our experimental setup along with the flow configurations.
A bacterium swimming to the left (along the $+Z$ direction) is held
by a strongly focused IR laser ($\lambda$= 1024 $nm$). In the absence
of the flow, the bacterium is invariantly held by the tail of the
body. The thrust force and the trapping force are balanced and the
bacterium is stationary with respect to the trap. The trapping remains
stable when a uniform flow in the $+Z$ direction is applied $(U>0)$
\cite{com1:}. The bacterium can also be trapped at the head of the
body, but the flow field must be reversed $(U<0)$.

To measure the trapping force and the position of the trapped cell
tip, the transmitted IR beam was refocused by a high numerical aperture
condenser (N.A. 1.5) and projected onto a two-dimensional position
sensitive detector (Pacific Silicon Sensor Inc., DL100-7PCBA). The
position of the trapped cell tip with respect to the center of the
trap is monitored by a PC equipped with a National Instruments analog-digital
converter card (AT-MIO-16E-2). The conversion rate is 10 kHz at 12-bit
resolution. A non-flagellated bacterium was used to calibrate the
spring constant $k$ of the optical trap by measuring the position
of the trapped tip as a function of the flow $U$. For the laser intensity
(23 mW) used in this experiment, $k$ = $5.7\times10^{-6}N/m$. A
brief description of the calibration process is presented in Materials
and Methods.

\begin{figure}

\centerline{
\includegraphics[width=6.5cm]{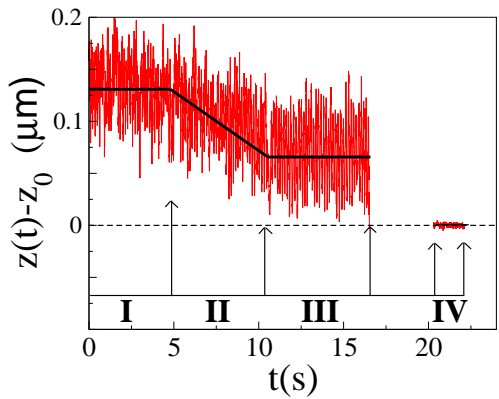}
}

\caption{A typical experimental run for a swimming bacterium held in the optical
trap. An uniform flow $U=10~\mu m/s$ is established in regime I.
The flow $U$ is decreased to zero linearly in regime II. The flow
$U$ remains zero in regime III. The laser is switched off momentarily
to let the bacterium escape and the undeflected laser beam position
is recorded in regime IV. Solid lines depict linear fits to each regimes.\label{cap:zvst}}

\end{figure}

Figure \ref{cap:zvst} displays an example of the time trace $z(t)$,
the longitudinal displacement along the swimming direction of a trapped
bacterium. We observed large oscillations overlying a systematic variation
of $z(t)$ as the external flow is changed. These oscillations result
from wobbling of the cell body in response to the rotation of the
flagella bundle \cite{rowe,berg:rapid-rotation}. The trapped bacterium
was perturbed by the following sequence of events: In regime I, the
bacterium is subject to a uniform flow $U=+10~\mu m/s$. The bacterium
maintains an average position away from the center of the trap. In
regime II, $U$ is linearly reduced to zero in $5~s.$ The average
bacterium position shifts systematically toward the center of the
trap. In regime III, $U$ is maintained at zero for $5~s$, and the
average position of the bacterium relative to the trap is again constant.
Finally in regime IV, the bacterium is released. The position of the
undeflected beam in regime IV is taken to be $z_{o}$, the center
of the optical trap. From regime II we obtain the net translational
drag coefficient $A+A_{0}$ = $k\Delta z/\Delta U$, and in regime
III we obtain $F_{thrust}$, since $F_{trap}=F_{thrust}$ when $U=0$.
We checked that this measurement was reproducible by returning the
flow to $U=10~\mu m/s$ rather than releasing the bacterium after
regime III. The bacterium returned to within a few percent of its
initial average position.

\begin{figure}

\centerline{
\includegraphics[width=8.5cm]{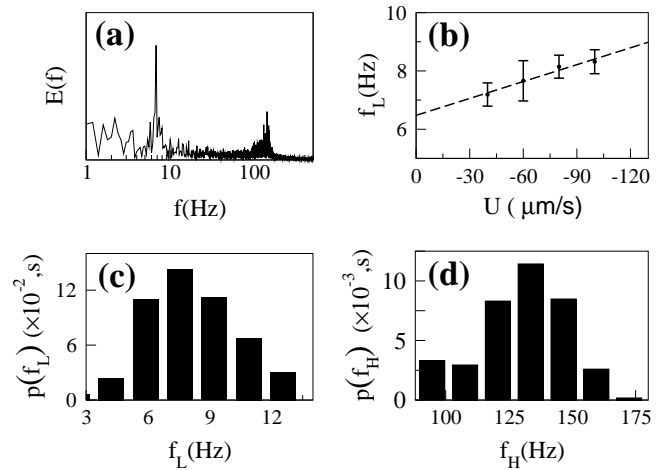}
}

\caption{(a) Power spectrum of $x(t)$ shows peaks corresponding to $f_{L}$
and \textbf{$f_{H}$}. (b) The variation of the rotation frequency
of the cell body $f_{L}$ as a function of flow speed $U$. The linear
dependence is consistent with the propulsion matrix formulation. (c)
and (d) delineate the PDFs of $f_{L}$ and $f_{H}$, respectively.
Error bars are standard errors of the mean unless otherwise noted.
\label{cap:powerspectrum}}

\end{figure}

We extracted the angular velocities $\Omega$ and $\omega$ using
a Fourier analysis of the time trace $x(t)$ of the transverse position
of the cell body. This transverse signal shows more pronounced oscillations
than $z(t)$. Figure \ref{cap:powerspectrum}(a) displays a sample
power spectrum $E(f)$ for a short time trace of $5~s$ when $U=0$
(regime III). The power spectrum has two strong peaks at $f_{L}\simeq7~Hz$
and $f_{H}\simeq130~Hz$, respectively. These two frequencies can
be associated with the angular velocities of the cell-body $\Omega=2\pi f_{L}$
and of the flagellum bundle $\omega=-2\pi f_{H}$ \cite{rowe}. Averaging
over 250 bacteria, we found $\bar{f}_{L}=(8.0\pm0.2)~Hz$ and $\bar{f}_{H}=(125\pm2)~Hz$,
where the $\pm$ are standard errors of the mean. However, as shown
in Fig. \ref{cap:powerspectrum}(c-d) there is considerable variation
of $f_{L}$ and $f_{H}$ between individual bacteria. The standard
deviations $\sigma_{f_{L}}=2.4$ Hz and $\sigma_{f_{H}}=27~Hz$ are
respectively 20 and 30 \% in the mean values.

To test the basic physics implied by the propulsion matrix, we measured
the dependence of $\bar{f}_{L}$ and $\bar{f}_{H}$ on $U$ for an
additional 40 bacteria which were subjected to flow speeds of $U=-40,-60,-80$
and $-100~\mu m/s$. Figure \ref{cap:powerspectrum}(b) shows that
the average frequency $\bar{f}_{L}$ increases linearly with $-U$
and the result is in good agreement with Eq. \textbf{\ref{eq:basicb}},
as predicted by the propulsion matrix formulation. Within the noise
of the measurement, no systematic change in $\bar{f}_{H}$ was detected.
This is expected since at low-loads the molecular motor is known to
rotate at a constant angular speed independent of the load \cite{berg:torque-speed}.

\begin{figure*}
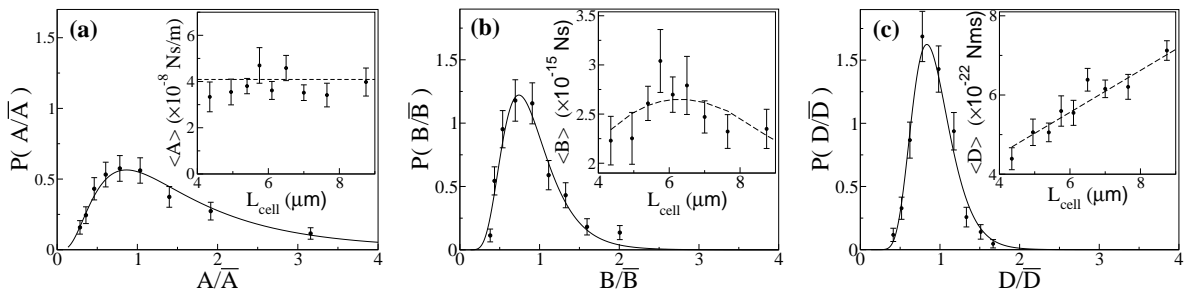


\centerline{
\includegraphics[width=5cm]{4a}~~
\includegraphics[width=5cm]{4b}~~
\includegraphics[width=5cm]{4c}
}

\caption{The PDFs of (a) $A/\bar{A}$, (b) $B/\bar{B}$, and (c) $D/\bar{D}$.
The insets shows the length $L_{cell}$ dependence of (a) $\langle A\rangle$,
(b) $\langle B\rangle$, and (c) $\langle D\rangle$. The solid lines
in the main figures are fits to log-normal distributions. \label{cap:ABD}}
\end{figure*}

To complete our determination of the propulsion matrix, the semi-minor
axis $a$ and length $L_{cell}=2b$ of the bacteria were measured
directly by video microscopy while immobilized in the trap. This allows
us to calculate the drag coefficients $A_{0}$ and $D_{0}$ for the
cell body. From the time trace $z(t)$, $A$ and $B$ are calculated
by $A$ = $k\Delta z/\Delta U-A_{0}$ and $B=F_{thrust}/\omega$ when
$U=0$. Finally, the measurements of the angular velocities gives
$D=-D_{0}\omega/\Omega.$ The matrix elements averaged over a population
of 250 bacteria are $\bar{A}=(3.8\pm0.2)\times10^{-8}~N~s/m$, $\bar{B}=(4.0\pm0.1)\times10^{-16}~N~s$,
and $\bar{D}=(5.9\pm0.1)\times10^{-22}N~s~m$. The translational drag
coefficient of the flagella is approximately twice that of the cell
body ($\bar{A_{0}}=1.7\times10^{-8}N~s/m$). Therefore a significant
portion of drag is due to the flagella. On the other hand, the rotational
drag of the flagella $\bar{D}$ is much smaller than that of the cell
body $(\bar{D}_{0}=8.8\times10^{-21}N~s~m)$.

It is instructive to use the measured propulsion matrix to extract
physical parameters that are relevant to flagellar bundles. The resistive
force theory for the helix coil contains four independent parameters:
the pitch $\lambda$, the pitch angle $\Psi$, the length of the helix
$\ell$, and the radius $r$ of the filament, assuming that the viscosity
of the fluid ($\eta=10^{-3}\, Pa\, s$) is known. The three matrix
elements $A$, $B$, and $D$ in Eqs. \textbf{\ref{eq:coefficient}}
however are not sufficient to predict all the four geometric parameters.
To make progress, we assumed that the the pitch angle $\Psi$ is $41^{o}$,
as determined by Turner et al. using fluorescently labeled E. coli
cells \cite{berg:fluorescent}. This angle also turns out to be remarkably
close to the optimal angle ($42^{o}$) that maximizes the propulsion
efficiency of an ideal helix \cite{Lighthill:BiofluidDynamics}. Using
$\beta=\cos^{2}(41^{o})=0.57$, Eqs. \textbf{\ref{eq:coefficient}}
predict $\gamma_{k}=0.84$, $\lambda=0.9\,\mu m$, $r=23\, nm$, and
$\ell=6.2\,\mu m$. These values are comparable to the fluorescent
measurements of Turner et al. who found $\bar{\ell}=7\,\mu m$, and
$\bar{\lambda}\simeq1\,\mu m$ for the curly flagella and $2.2\,\mu m$
for the normal ones \cite{berg:fluorescent}. We can estimate the
average number of flagella using $r\approx\sqrt{\bar{N}}~r_{o}$,
which gives $\bar{N}\approx r^{2}/r_{o}^{2}\approx5.3$. This is slightly
greater than the $\bar{N}\approx3.3$ found by Turner et al. The difference
may be expected because the bacteria used in their experiment are
shorter than the ones we studied; longer bacteria usually have more
flagella.

All important dynamical quantities can be obtained from our measurements.
For example, the average thrust for $U=0$ is $\bar{F}_{thrust}=\bar{B}~\bar{\omega}=0.31~pN$.
The average swimming speed is $\bar{V}_{swim}=\bar{B}~\bar{\omega}/\bar{(A}_{0}+\bar{A)}=6~\mu m/s$,
which should be compared with $V_{swim}\approx10~\mu m/s$ we obtained
directly by video microscopy. The difference may be due to correlations
between $A$, $B$ and $D$; both $B$ and $D$ grows with $A$ on
average. Similarly the average torque $\bar{N}_{fl}=\bar{D}~(|\bar{\omega}-\bar{\Omega}|)=4.9\times10^{-19}N~m$
is surprisingly close to that found for \emph{Streptococcus} \cite{berg:torque-speed}.

We observed that the propulsion matrix elements vary greatly among
individual bacteria even though our bacteria are from a single colony.
Figure \ref{cap:ABD} shows the probability distribution functions
(PDF) of the scaled quantities $A/\bar{A}$, $B/\bar{B}$, and $D/\bar{D}$.
The standard deviations $\sigma$ are comparable to the mean values
with $\sigma_{A}/\bar{A}=0.7$, $\sigma_{B}/\bar{B}=0.5$ and $\sigma_{D}/\bar{D}=0.3$.
A conspicuous feature of all the PDFs is their broad tails, particularly
for the linear drag coefficient $A$. This might be an indication
of either significant structural heterogeneities in the flagellar
bundles of individual cells or of changes in conformations of the
flagellar bundle with time. As is often the case in biological systems,
the PDFs can be roughly fitted to log-normal distribution functions,
which are plotted as solid lines in the figure.

\begin{figure*}
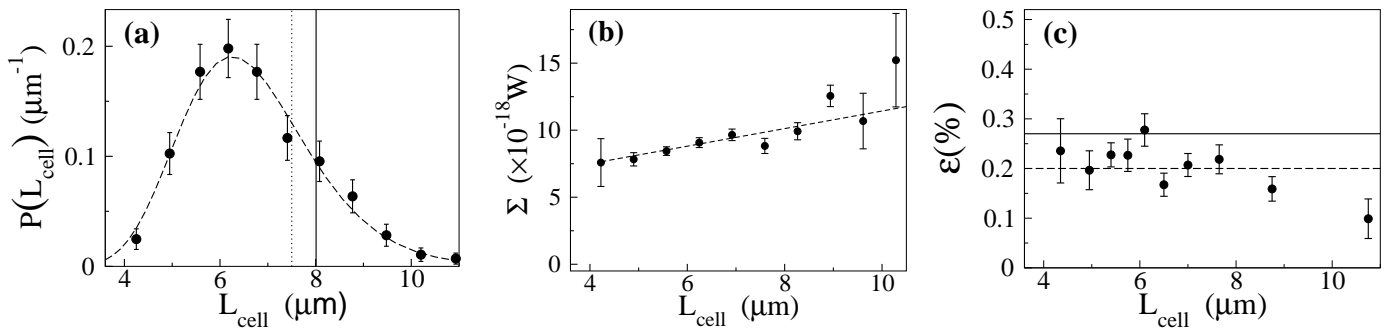


\centerline{
\includegraphics[height=4.25cm,clip]{5a}~~~
\includegraphics[height=4.25cm,clip]{5b}~~
\includegraphics[height=4.25cm,clip]{5c}
}

\caption{(a) The PDF of the bacterial cell length $L_{cell}$. The dashed
line is a fit to the log-normal distribution. The dotted and solid
vertical lines are, respectively, the cell lengths at which we first
observed a septal ring and where cell division occurred. (b) The flagellar
power output $\langle\Sigma\rangle$ as a function of $L_{cell}$.
The dashed line is a linear fit. (c) The propulsion efficiency $\langle\varepsilon\rangle$
as a function of $L_{cell}$. The dotted horizontal line marks the
mean efficiency 0.2\% of the entire population. The solid horizontal
line is the maximum efficiency $\bar{\varepsilon}_{max}$ when the
flagellum length is optimized. See text for details. \label{cap:efficiency}}

\end{figure*}

Part of the variation in $A,$ $B$ and $D$ must arise due to bacteria
being in different stages of their growth cycle during the measurements.
This is particularly the case in the middle log-phase of a growing
culture, where the bacterial size is highly varied. For ease of trapping,
very long and very short bacteria were excluded from the measurements
and the middle-sized bacteria ($4-12\,\mu m$), which comprised about
$95\%$ of the population, was chosen. The bacterial cell-length distribution
of this selected population is plotted in Fig. \ref{cap:efficiency}(a).
The figure also shows the cell length at which we first observed septal
rings (dotted line) and the length at which cells divide (solid line).
We used the bacterial length $L_{cell}$ as a measure of the bacterium's
physiological state and determine the propulsive matrix elements as
a function of $L_{cell}$. To determine the length dependence of the
coefficients $A$, $B$, and $D$, we calculated the averaged values
$\langle A\rangle$, $\langle B\rangle$ and $\langle D\rangle$ for
bacteria of similar length. The result are presented in the insets
of Fig. \ref{cap:ABD}(a-c). The linear drag coefficient $\langle A\rangle$
has no clear size dependence but $\langle B\rangle$ is peaked at
$L_{cell}\approx6~\mu m$, which coincides with the peak of the size
distribution. On the other hand, Fig. \ref{cap:ABD}(c) shows that
the rotational drag coefficient $\langle D\rangle$ of the propeller
grows linearly with the cell-body length $L_{cell}$.

These size dependencies allow us to draw conclusions about the structure
of flagellar bundles at different stages of cell growth. Inspection
of Eqs.\ {}\textbf{\ref{eq:coefficient}} shows that the matrix elements
scale with the pitch $\lambda$ according to $A\propto\lambda^{0}$,
$B\propto\lambda^{1}$, and $D\propto\lambda^{2}$. The fact that
we find that $B$ and $D$ depend on $L_{cell}$ but $A$ does not,
implies that the primary $L_{cell}$ dependence is in pitch $\lambda$.
The pitch angle $\Psi$ and the flagella length $\ell$ are approximately
constant independent of $L_{cell}$. Likewise, $\gamma_{k}$ depends
logarithmically on $\lambda$ and so is only very weakly dependent
on $L_{cell}$. Since our measurements show a linear relation between
$D$ and $L_{cell}$, it suggests that $\lambda^{2}$ grows linearly
with $L_{cell}$. A possible scenario for this is that more and more
flagella are incorporated into the bundle as the bacteria cell body
grows and this causes the $\lambda^{2}$ to grow in agreement with
Fig.\ \ref{cap:ABD}(c). From the shortest to the longest bacterial
body size ($L_{cell}$), we found that the fractional change $\delta\lambda/\lambda$
is about $16\%$, which may be discernible in carefully conducted
observations using fluorescently labeled bacteria.

We next turn our attention to the power and propulsive efficiency
of the swimming bacteria. The average power output of the flagellar
motors is $\bar{\Sigma}=D_{0}\Omega(\omega-\Omega)=0.4~pW$. The power
used to turn the cell body is $D_{0}\Omega^{2}\approx0.02~pW$ while
the actual propulsive power is another factor of ten smaller with
$A_{0}V_{swim}^{2}\approx0.0017~pW$. Therefore $5\%$ of the rotary
power is used to rotate the cell body, and only $0.5\%$ is used to
push the bacteria forward. Figure \ref{cap:efficiency}(b) shows the
average motor power as a function of bacterial length $L_{cell}$.
The power increases gradually with $L_{cell}$, which is consistent
with the above discussion that the number of flagella $N$ and the
associated motors increase with $L_{cell}$. The propulsion efficiency
$\varepsilon$, defined as the ratio of the propulsive output power
to the rotary input power, can be related to the propulsion matrix
elements \cite{purcell}:
\begin{widetext}
\begin{eqnarray}
	\varepsilon & \equiv & \frac{A_{0}v^{2}}{N_{fl}(\omega-\Omega)}
		= \frac{A_{0}D_{0}B^{2}}{[(A_{0}+A)D-B^{2}][(A_{0}+A)(D_{0}+D)-B^{2}]}
		\approx \frac{A_{0}B^{2}}{(A_{0}+A)^{2}D}
	\label{eq:efficiency}
\end{eqnarray}
\end{widetext}
 Here we used $B^{2}\ll(A_{0}+A)D$ and $D_{0}\gg D$ to obtain the
approximate form. These assumptions are met on average but does not
always hold for a particular bacterium. Therefore we use the full
form to calculate the efficiency. Figure \ref{cap:efficiency}(c)
shows that the efficiency as a function of bacterial size is constant
up to the cell division length ($L_{cell}=8~\mu m$). The average
efficiency $\bar{\varepsilon}\approx0.2\%$ \cite{com6:} is surprisingly
close to that of sedimenting helices in a silicon oil, which were
tested as model flagella by Purcell \cite{purcell}. We can also ask,
for a given $A_{0}$, what is the maximum efficiency attainable by
the bacterium as a function of the length of the flagella. Assume
that at some characteristic length $\ell_{p}$, the propulsive coefficients
of the flagellum are $A_{p}$, $B_{p}$ and $D_{p}$. Neglecting logarithmic
corrections and assuming the width of the flagellar bundle is constant,
these coefficients should grow linearly with flagella length $\ell$
so that $A\approx\kappa A_{p}$, $B\approx\kappa B_{p}$ and $D\approx\kappa D_{p}$
where $\kappa=\ell/\ell_{p}$. This assumption is consistent with
Eqs. \textbf{\ref{eq:coefficient}}. Substituting for $A$, $B$ and
$D$ into our approximate expression for $\varepsilon$ (in Eq. \textbf{\ref{eq:efficiency}})
we find that the maximum efficiency occurs when $A=A_{0}$ and that
$\varepsilon_{max}\approx B_{p}^{2}/(4A_{p}D_{p})$ which depends
only on the shape of the propeller. The same result was obtained by
Purcell when he maximized $\varepsilon$ by assuming that all propeller
dimensions (not just the length) scaled with $\kappa$ \cite{purcell}.
In our experiments, we find that $A\approx2A_{0}$ and so flagella
are twice as long as that required to maximize its propulsive efficiency.
However, the peak in $\varepsilon$ as a function of $\kappa$ is
fairly broad and the observed efficiency is about $75\%$ of the maximum
efficiency as shown by the solid and dashed lines in Fig. \ref{cap:efficiency}c.
The broadness of the peak may be why the propulsive efficiency is
approximately constant throughout the bacterial cell division cycle.
The experimental propulsive efficiencies are consistent with the small
$\varepsilon_{max}$ of between $0.3\%-0.8\%$ measured for helical
propellers \cite{purcell}. It is smaller still than the 1 or 2\%
predicted theoretically for a helical propeller \cite{Lighthill:BiofluidDynamics,childress}.

In summary, bacterial propulsion in a uniform stream is investigated
with the help of optical tweezers, which allow the thrust force $F_{thrust}$
to be directly measured as a function of imposed flow. For a free
swimming bacterium, $F_{thrust}$ precisely balances the viscous drag
of the cell body $A_{0}v$ and of the flagellar bundle $Av$. The
contribution to the drag by $Av$ is twice as large as $A_{o}v$ but
is difficult to determine without our direct force measurements. We
also showed that the propulsion matrix description proposed by Purcell
gives an adequate description of bacterial propulsion over a physiological
range of velocities. In retrospect, the validity of the propulsion
matrix, or for that matter the resistive force theory itself, is not
self evident for real micro-organisms because of possible deformations
of flagellar bundles due to hydrodynamic stresses induced by swimming
or by the flow \cite{childress}.

We have determined all elements of the propulsion matrix and used
the resistive force calculations for a helical coil to estimate microscopic
properties of flagella \cite{Lighthill:BiofluidDynamics}. The results
were consistent with earlier measurements even though the resistive
force calculations neglect the effect of long-range hydrodynamics
interactions between different parts of the flagella and with the
cell body. Using the propulsion matrix, we have determined all dynamic
quantities related to bacterial swimming and their dependence on the
size of the cell body. In particular, we found that the propulsive
efficiency $\varepsilon$, defined as the ratio of the propulsive
power output to the rotary power input, is $\sim0.2\%$ and is nearly
independent of $L_{cell}$. The measured $\varepsilon$ is consistent
with the experiments on rigid helical propellers \cite{purcell} and
close to the maximum efficiency for the given size of the cell body
and the shape of the flagellar bundle. A conspicuous finding of our
measurements is that all the matrix elements are broadly distributed
despite the fact that all bacteria started from a single colony. The
ubiquity of such broad distributions in biological systems is significant
and begs further systematic study.

\ \\

\noindent \textbf{MATERIALS AND METHODS}

\noindent Sample preparation: We followed standard growth conditions
for culturing bacteria E. coli strains, RP5231 and YK4516. RP5231
is a smooth swimming strain because two of its chemotactic genes,
CheY and CheZ, were deleted. A single colony was picked from a fresh
agar plate and grown to saturation overnight in the LB medium (peptone
4 g, yeast 2 g, NaCl 1 g, 1M NaOH 0.4 ml; for 400 ml of media). The
culture was maintained at 30 $^{o}C$ and was shaken continuously
at 200 rpm. The overnight sample was diluted 1:100 in fresh LB medium
and grown to the middle log phase for 3 hours.

To calibrate the spring constant $k$ of the optical trap, we used
non-flagellated bacteria strain YK4516. A uniform flow $U$ was applied
and the shift in the centroid of the transmitted IR laser beam was
recorded by the position detector. For an ellipsoid body, the translational
drag coefficient $A_{o}$ is known and the spring constant is obtained
using $k=A_{o}\Delta U/\Delta z$. The noise in the output of the
optical trap was $0.1\, nm/\sqrt{Hz}$ (for $z_{rms}=5.2\, nm$ and
sampling rate $10KHz$). 
\\

\bf
\noindent
ACKNOWLEDGEMENTS
\rm

We would like to thank Emily Chapman, Roger Hendrix, and Bob Duda
for helpful discussions and technical assistance. This research is
supported by the National Science Foundation under Grant no. DMR-0242284.


\begin{thebibliography}{10}
\bibitem{Berg:PoissonMotor}Samuel, A. D.~T. \& Berg, H.~C. \newblock (1996) \emph{Biophys.
J.} \textbf{71}, 918--923.
\bibitem{berry:steps}Sowa, Y., Rowe, A.~D., Leake, M.~C., Yakushi, T., Homma, M., Ishijima,
A., \& Berry, R.~M. \newblock (2005) \emph{Nature} \textbf{437},
916--919.
\bibitem{taylor:wavingtails}Taylor, G.~I. \newblock (1952) \emph{Proc. Roy. Soc. London A} \textbf{211},
225--239.
\bibitem{Lighthill:BiofluidDynamics}Lighthill, J. \newblock (1989) \emph{Mathematical Biofluiddynamics}.
\newblock (SIAM, Philadelphia), 3$^{rd}$ edition, pp. 45--92.
\bibitem{childress}Childress, S. \newblock (1981) \emph{Mechanics of swimming and flying},
Cambridge Studies in Mathematical Biology. \newblock (Cambridge University
Press, New York), 1$^{st}$ edition, pp. 34--60.
\bibitem{berg:proton}Meister, M., Lowe, G., \& Berg, H.~C. \newblock (1987) \emph{Cell}
\textbf{49}, 643--650.
\bibitem{berg:fluorescent}Turner, L., Ryu, W.~M., \& Berg, H.~C. \newblock (2000) \emph{J.
Bacteriol.} \textbf{182}, 2793--2801.
\bibitem{magariyama}Magariyama, Y., Sugiyama, S. \& Kudo, S. \newblock (2001) \emph{FEMS
Microbiol. Lett.} \textbf{199}, 125--129.
\bibitem{purcell}Purcell, E.~M. \newblock (1997) \emph{Proc. Natl. Acad. Sci. USA}
\textbf{9}, 11307--11311.
\bibitem{happel:lowreynolds}Happel, J. \& Brenner, H. \newblock (1965) \emph{Low Reynolds Number
Hydrodynamics with Special Applications to Particulate Media}. \newblock (Prentice-Hall,
Englewood Cliffs), 1$^{st}$ edition, pp. 173--183.
\bibitem{berg:RandomWalk}Berg, H.~C. \newblock (1993) \emph{Random Walks in Biology}. \newblock (Princeton
University Press, New Jersey), 1$^{st}$ edition, pp. 57--84.
\bibitem{bustamante:trap}Wuite, G.~J., Davenport, R.~J., Rappaport, A., \& Bustamante, C.
\newblock (2000) \emph{Biophys. J.} \textbf{79}, 1155--1167.
\bibitem{Neuman:opticalreview}Neuman, K. \& Block, S. \newblock (2004) \emph{Rev. Sci. Inst.} \textbf{75},
2787--2809.
\bibitem{com7:} \newblock We found that it was difficult to stably trap and measure
the force on the bacteria, when it is swimming perpendicular to the
optical trap, in the interior of the fluid. The presence of the surface
may have some effect on our measurements but does not change our qualitative
conclusions.
\bibitem{com1:} \newblock We introduced the uniform flow $U$ by translating the
sample chamber using a Newport DC motor controller (Model 855) and
two motorized translation stages. By controlling the velocities of
two oppositely moving stages, a broad range of $U$ (from 0 to 100
$\mu$m/s) can be achieved.
\bibitem{rowe}Rowe, A.~D., Leake M.~C., Morgan, H., \& Berry, R.~M. \newblock (2003)
\emph{J. of Mod. Opt.} \textbf{50}, 1539--1554.
\bibitem{berg:rapid-rotation}Lowe, G., Meister, M., \& Berg, H.~C. \newblock (1987) \emph{Nature}
\textbf{325}, 637--640.
\bibitem{berg:torque-speed}Chen, X. \& Berg, H.~C. \newblock (2000) \emph{Biophys. J.} \textbf{78},
1036--1041.
\bibitem{com6:} \newblock Note that due to the correlation of $A$, $B$ and $D$,
substituting the average values $\bar{A}$, $\bar{B}$, $\bar{D}$
into Eq. \textbf{\ref{cap:efficiency}} only gives an efficiency of
$0.16\%$ instead of $0.2\%$. \end{thebibliography}
\end{document}